\documentclass[lettersize,journal]{IEEEtran}
\usepackage{amsmath,amsfonts}
\usepackage{algorithmic}
\usepackage{algorithm}
\usepackage{array}
\usepackage{textcomp}
\usepackage{stfloats}
\usepackage{url}
\usepackage{verbatim}
\usepackage{graphicx}
\usepackage{cite}
\usepackage{subcaption}
\usepackage{mathtools}
\usepackage{graphicx}

\newcommand{\Vr}{\mathbf{V_R}}
\newcommand{\Vi}{\mathbf{V_R}}
\newcommand{\Ir}{\mathbf{I_R}}
\newcommand{\Ii}{\mathbf{I_R}}
\newcommand{\gr}{\mathbf{g_{nn}^R}}
\newcommand{\gi}{\mathbf{g_{nn}^I}}

\begin{document}

\title{Integrating Forecasting Models Within Steady-State Analysis and Optimization}

\author{Aayushya Agarwal, Larry Pileggi
\thanks{A.A., L.P. are with the Department of Electrical and Computer Engineering at Carnegie Mellon University, Pittsburgh, PA, 15213 (email: {aayushya, pileggi}@andrew.cmu.edu.}
}

\markboth{Journal of \LaTeX\ Class Files,~Vol.~14, No.~8, August~2021}%
{Shell \MakeLowercase{\textit{et al.}}: A Sample Article Using IEEEtran.cls for IEEE Journals}


\maketitle

\begin{abstract}
Extreme weather variations and the increasing unpredictability of load behavior make it difficult to determine power grid dispatches that are robust to uncertainties. While machine learning (ML) methods have improved the ability to model uncertainty caused by loads and renewables, accurately integrating these forecasts and their sensitivities into steady-state analyses and decision-making strategies remains an open challenge. Toward this goal, we present a generalized methodology that seamlessly embeds ML-based forecasting engines within physics-based power flow and grid optimization tools. By coupling physics-based grid modeling with black-box ML methods, we accurately capture the behavior and sensitivity of loads and weather events by directly integrating the inputs and outputs of trained ML forecasting models into the numerical methods of power flow and grid optimization. Without fitting surrogate load models, our approach obtains the sensitivities directly from data to accurately predict the response of forecasted devices to changes in the grid. Our approach combines the sensitivities of forecasted devices attained via backpropagation and the sensitivities of physics-defined grid devices. We demonstrate the efficacy of our method by showcasing improvements in sensitivity calculations and leveraging them to design a robust power dispatch that improves grid reliability under stochastic weather events. Our approach enables the computation of system sensitivities to exogenous factors which supports broader analyses that improve grid reliability in the presence of load variability and extreme weather conditions. 
\end{abstract}


\section{Introduction}
\IEEEPARstart{T}{he} rise of renewable energy sources, the growth in AI-driven demand, and the increasing variability in weather conditions have introduced significant unpredictability into the electrical grid. Unlike traditional loads and generators, the power consumed or produced by these devices is heavily influenced by exogenous factors such as weather conditions, time-of-day, and user behavior.

The stochastic nature of these devices poses a challenge for maintaining real-time grid stability and long-term reliability. To combat this, grid operators rely on forecasting methods to model the uncertainty in power generation and consumption. These forecasting methods use historical data to train statistical and machine learning models that predict how individual devices behave in response to non-physical factors including weather or user patterns.

The outputs are then used to parameterize surrogate models, such as PQ models, that are then integrated into optimization engines for designing dispatches and planning decisions. However, abstracting the outputs of forecasting methods to surrogate PQ models discards valuable information on the impact of system wide performances that can be used to design robust optimization methods. The key challenge, then, is how to integrate machine learning forecasting engines directly into physics-based solvers to retain the non-physical behavior captured by ML models while leveraging the accuracy of physics-based optimization.

In this work, we introduce a framework that directly integrates forecasting models into power flow and optimal power flow (OPF) solvers. The advantage of this approach is that it allows us to compute the impact of variables influencing each forecasted device, such as weather or user demand patterns, on system performance by combining the physics-based sensitivities of deterministic grid components with the data-driven sensitivities from machine learning forecasts. This enables better decision-making under uncertainty and can be used within an optimization engine to develop grid-scale efforts such as dispatches robust to large weather fluctuations, secondary control mechanisms for unpredictable load behaviors, and planning strategies that are robust to environmental changes.

We demonstrate the effectiveness of our methodology by first deriving system-wide sensitivities to non-physical features such as weather conditions and user behavior. These sensitivities offer operators valuable, real-time insight into how exogenous factors impact grid performance. We then show how these sensitivities can be utilized to create new optimization analyses accounting for weather patterns and user behaviors. We demonstrate this approach by developing a dispatch algorithm that is robust to large weather fluctuations, using the sensitivities generated by our approach.

Our approach opens the door to broader studies that explore new types of sensitivities critical to future grid planning. As load variations and extreme weather events become more common, our approach provides a foundation to study the impact of evolving weather patterns on system performances. This capability is especially valuable for guiding investment, resilience planning, and new optimization efforts in increasingly complex and dynamic power systems.

\section{Previous Work}
Grid operators rely on forecasting engines to predict the power generated or consumed by individual devices in the grid, including renewable energy sources (such as wind turbines and photovoltaic systems) and electrical loads (such as households and sub-stations). Accurate forecasting is essential for reliable grid operation, efficient energy trading, and the integration of renewable energy sources \cite{hong2015weather}\cite{weron2014electricity}. Forecasting methods are broadly categorized into statistical methods and machine learning (ML) methods.

Statistical models have been used for power system forecasting due to their interpretability and lower computational requirements. Autoregressive Integrated Moving Average (ARIMA) models have been widely applied in short-term load and wind power forecasting due to their effectiveness in modeling time-series data with trends and seasonality \cite{hyndman2018forecasting}\cite{taylor2003short}. Exponential smoothing methods, including the Holt-Winters technique, are used for univariate time series and perform well for seasonal load profiles \cite{hippert2002neural}. State Space Models and Kalman Filters have been used to update predictions in real-time, making them useful for applications requiring adaptive or online forecasting \cite{arora2021remodelling}. While statistical models are generally simple and require fewer data preprocessing steps, they often struggle in capturing the nonlinear and high-dimensional relationships inherent in power systems, particularly under high penetration of distributed and renewable energy resources, which continues to increase.

Machine learning (ML) methods have gained attention due to their capacity to model nonlinear relationships by training on large datasets. These methods are well-suited for forecasting devices with high variability and uncertainty, such as renewable power generation. Artificial Neural Networks (ANNs) have been widely applied to learn patterns of load, wind, and solar power forecasting from historical data \cite{wazirali2023state}\cite{eskandari2021convolutional}. Support Vector Regression (SVR) has been shown to perform well in both short- and medium-term forecasting by handling nonlinear relationships and avoiding overfitting \cite{wu2021support}. Ensemble methods, such as Random Forests (RF) and Gradient Boosting Machines (GBM), combine multiple ML models to reduce training variance and improve generalization across network conditions. These models have shown superior performance in several energy forecasting studies \cite{fan2022applications}\cite{matrenin2022medium}\cite{sobolewski2023gradient}. Deep Learning approaches, especially Recurrent Neural Networks (RNNs) and Long Short-Term Memory (LSTM) networks, have been used to model temporal dependencies and handle sequential data effectively. LSTM networks are particularly adept at capturing long-term dependencies in time-series data and have been successfully applied to both wind and solar forecasting \cite{dubey2021study}\cite{huang2022time}\cite{aseeri2023effective}.

While forecasting methods have become increasingly accurate in predicting power generation or consumption, they are embedded within power flow and optimization engines via surrogate models. A common approach involves using forecasted data to fit parameters for standard surrogate models, such as PQ (active/reactive power) or composite models that approximate device behavior under varying electrical conditions. Surrogate models simplify the representation of forecasted devices into explicit forms for inclusion in grid constraints. However, these models assume a voltage-current sensitivity that may not accurately reflect real-world operational characteristics. This can lead to simulations that diverge from the real behavior of the grid, providing grid operators with unrealistic results. Additionally, abstracting forecasted outputs into surrogate functions often discards critical information about how exogenous factors, such as weather and time-of-day, influence device behavior.

Our contribution focuses on the integration of ML-based forecasting methods into power flow and optimization engines to improve the accuracy of sensitivity models and better capture the influence of non-physical factors on system metrics. Embedding these data-driven insights into the optimization process enables more informed and adaptive decision-making that will improve power modeling accuracy and grid resilience.

\section{Integrating Forecasting Models into Power Flow and Optimization}

Prior methods integrate forecasts into decision-making optimization strategies via surrogate PQ and composite models to represent a forecasted device behavior. Our approach directly integrates the forecasting model into the numerical engines of power flow and grid optimization to achieve the following benefits:

\begin{enumerate}
    \item Avoid assumptions in surrogate models for more accurate voltage sensitivity and simulations
    \item Extract sensitivities of each forecasted device caused by exogenous factors
    \item Use the sensitivities of individual forecasted devices to measure impact of exogenous factors on system-wide performance metrics
\end{enumerate}
To achieve these advantages, we first develop a forecasting model that uses state-variables of the grid as an input. This allows us to bypass the surrogate model and integrate forecasting models within power flow and optimization engines.

\subsection{Building I-V Forecasting Models}

To integrate forecasting methods directly into the grid optimization, we introduce a current-voltage (I-V) forecasting model that predicts the complex current drawn by a device based on both non-physical features (such as weather conditions and user behavior) and state variables of the grid (bus voltage). The I-V forecasting model, $g_{nn_i } (\cdot)$, is a machine-learning (ML) model that predicts the real and imaginary currents of a forecasted device on bus $i$ as:
\begin{equation}
    \begin{bmatrix}
        I_R \\ I_I
    \end{bmatrix} = g_{nn_i}(w, V_R, V_I, u),
    \label{eq:I-V_forecast_eq}
\end{equation}
where $I_R,I_I$ are the real and imaginary components of the device current, and $V_R,V_I$ are the real and imaginary bus voltages. The vector $u$ captures exogenous features such as weather and time-of-day, which are also inputs in traditional forecasting methods \cite{wazirali2023state}. The forecasting model is parameterized by a vector of weights, $w$, that is learned from training.

The I-V forecasting model differs from traditional models that use only exogenous features (such as time of day and temperature) to predict active and reactive power (P and Q). Including the bus voltage as an input allows us to directly learn the current-voltage sensitivity from data as opposed to assuming it via surrogate PQ models. 
Training the proposed I-V forecasting model follows a supervised learning approach where the objective is to minimize the mean squared error between predicted and measured outputs. However, unlike traditional approaches, the inputs to the I-V forecasting model include both the non-physical features and the local bus voltages, while the outputs are the real and imaginary components of the device current. The supervised training optimizes the model’s parameters, $w$, as follows:
\begin{equation}
    \min_{w} \sum_{s\in \mathcal{S}} \left \| \begin{bmatrix}
        y_R^s\\ Y_I^s
    \end{bmatrix} - g_{nn_i}(w, V_R^s, V_I^s, u^s \right\|^2 
\end{equation}
where $s$ is a sample from the training dataset, $\mathcal{S}$. The model is trained on historical or simulated data ($y_R$ and $y_I$ represent measured real and imaginary device currents) that spans a span of voltage conditions, $V_R^s+jV_I^s$, and behavioral patterns, $u^s$. The forecasting model can follow prior methods including deep-neural networks, LSTMs and ensemble methods \cite{wazirali2023state, hippert2002neural}. Notably, by including voltage as an input feature, the I-V model learns the current-voltage characteristics of the forecasted device’s without relying on any prior assumptions. This provides an accurate simulation for studying the grid response under different operating scenarios. 

\subsection{Integrating I-V Forecast Models into Power Flow}
The I-V forecasting model bypasses the need for a surrogate model and can be directly integrated into power flow analyses. The current-voltage formulation of network constraints of the grid (described in \cite{bromberg2015equivalent}) is represented as:
\begin{align}
    &\sum_{j\in D_i} I_R^j(V_R^i,V_I^i)=0   \label{eq:kcl_eq1}
\\
   & \sum_{j\in D_i} I_I^j(V_R^i,V_I^i)=0 \quad \forall i\in\mathcal{B}
   \label{eq:kcl_eq2}
\end{align}
The constraints in the current-voltage formulation represent Kirchoff’s current law (KCL), in which the sum of real and imaginary currents (represented by $I_R^j$ and $I_I^j$) of the devices, $D_i$, at each bus, $i$, in the set of buses, $\mathcal{B}$,  must equal zero. The device currents are a function of the real and imaginary bus voltages, $V_R^i$ and $V_I^i$, respectively.

Using a current-voltage formulation of the network constraints, we can directly integrate the I-V forecast models defined in \eqref{eq:kcl_eq1}-\eqref{eq:kcl_eq2} as:
\begin{align}
    &\sum_{j\in D_i} I_R^j(V_R^i,V_I^i)+g_{nn_i}^R(\bar{w}_j, V_R, V_I,u_j)=0    \label{eq:kcl_forecast_eq1}
\\
   & \sum_{j\in D_i} I_I^j(V_R^i,V_I^i)+g_{nn_i}^I(\bar{w}_j, V_R, V_I,u_j)=0 \quad \forall i\in\mathcal{B}
   \label{eq:kcl_forecast_eq2}
\end{align}
where $g_{nn_i}^R$ and $g_{nn_i}^I$ represent the real and imaginary components of the forecast model in \eqref{eq:kcl_forecast_eq1}-\eqref{eq:kcl_forecast_eq2} on bus $i$. These models are evaluated with learned model weights, $\bar{w}_j$.

Equations \eqref{eq:kcl_forecast_eq1}-\eqref{eq:kcl_forecast_eq2} integrate the I-V forecasting models directly into the power flow constraints and bypass the need for a surrogate model as the current-voltage behavior is intrinsically modeled by $g_{nn_i}$.  This creates a synergy between physical models (defined by $I_R^i$  and $I_I^i$) and ML models (defined by $g_{nn_i}$). Solving the power flow embedded with I-V forecasting models in \eqref{eq:kcl_forecast_eq1}-\eqref{eq:kcl_forecast_eq2} requires the simultaneous integration of physical and forecast-based components into the underlying numerical method.

\paragraph{Solving Power Flow with I-V Forecast Models} We introduce an algorithm based on a Newton-Raphson solver that integrates the I-V forecasting models into the numerical engine. In our approach, each step of a Newton-Raphson method to solve \eqref{eq:kcl_forecast_eq1}-\eqref{eq:kcl_forecast_eq2} is expressed as follows:

\begin{multline}   
    \begin{bmatrix}
        \frac{d\Ir}{d\Vr}+\frac{d\gr}{d\Vr}, \frac{d\Ir}{d\Vi}+\frac{d\gr}{d\Vi} \\
        \frac{d\Ii}{d\Vr}+\frac{d\gi}{d\Vr}, \frac{d\Ii}{d\Vi}+\frac{d\gi}{d\Vi}
    \end{bmatrix} \begin{bmatrix}
        \Delta \Vr^k \\ \Delta \Vi^k
    \end{bmatrix} = \\ -\begin{bmatrix}
        \Ir(\Vr, \Vi, z) + \gr(\bar{w}, \Vr, \Vi, \mathbf{u}) \\ \Ii(\Vr, \Vi, z) + \gi(\bar{w}, \Vr, \Vi, \mathbf{u})
    \end{bmatrix}
\end{multline}

where $\Ir, \Ii$ are vectors of real and imaginary currents from all devices, and $\Vr, \Vi$ are vectors of real and imaginary bus voltages. Additionally, $\gr, \gi$ are vectors of real and imaginary currents from the trained I-V forecast models with trained parameters, $\bar{w}$.

In the Newton-Raphson step, we observe that the partials with respect to $\Ir$  and $\Ii$ are derived analytically. To integrate the trained ML models within the NR step, we must determine the partials, $\gr$ and $\gi$. This is computed using backpropagation workflow shown in Algorithm \ref{power_flow_alg}. 

The key benefit of integrating I-V forecast models into power flow is that we derive the sensitivities of the forecasted devices through data as opposed to relying on a surrogate model. As shown in the experiments, bypassing the assumptions of the surrogate model leads to more accurate simulations.

The key advantage of integrating I–V forecast models into power flow analysis is that device sensitivities are learned directly from data, rather than approximated through simplified surrogate models such as ZIP or constant PQ. These surrogate models rely on fixed assumptions that often fail to capture the nonlinear and context-dependent behavior of real-world devices under varying operating conditions. In contrast, data-driven I–V models accurately represent actual device responses, resulting in more precise power flow solutions. As demonstrated in our experiments, our approach improves simulation accuracy and enables more reliable gradient computations for optimization, control, and risk analysis, leading to more informed and adaptive grid operation.

\begin{algorithm}
    \caption{Backpropagation of I-V Forecasting Models Using PyTorch \cite{pytorch}}
    \label{power_flow_alg}
    \textbf{Input: } $g_{nn}$($\cdot$), $\bar{w},u,V_R,V_I$
    \begin{algorithmic}[1]
 
    \STATE{ $[I_R,I_I ]=g_{nn} (\bar{w},V_R,V_I,u)$)}
    \STATE{ $g_{nn}.$backward(retain\_graph=$True$)}
    \STATE{ $[\frac{d I_R}{d u}, \frac{d I_R}{d V_R}, \frac{d I_R}{d V_I}]$=$I_R$.grad.detach().clone()}
    \STATE{ $[\frac{d I_I}{d u}, \frac{d I_I}{d V_R}, \frac{d I_I}{d V_I}]$=$I_I$.grad.detach().clone()}
  
    \RETURN $[\frac{d I_R}{d u}, \frac{d I_R}{d V_R}, \frac{d I_R}{d V_I},\frac{d I_I}{d u}, \frac{d I_I}{d V_R}, \frac{d I_I}{d V_I} ]$
    \end{algorithmic}
    \end{algorithm}

\subsection{Integrating I-V Forecast Models into Optimization}
The I-V forecasting model can also be directly integrated into grid optimization problems to enhance the accuracy of dispatches and planning decisions. Consider a general optimization function:
\begin{subequations}
\label{eq:opf_eq}
    \begin{equation}
        \min_z f(z)
    \end{equation}
    \begin{equation}
       s.t. \sum_{i \in D_j} I_R^j (V_R^i,V_I^i,z) =0 
    \end{equation}
    \begin{equation}
       \sum_{i\in D_j} I_I^j (V_R^i,V_I^i,z) =0 \quad \forall i\in \mathcal{B} 
    \end{equation}
\end{subequations}
 $z$ represents a vector of decision variables such as active power of synchronous generators or line impedances. The objective function, $f(z)$, represents to various operational and planning problems including economic dispatch, transmission losses, and cost of ancillary services. The constraints of the optimization problem represent the network constraints formulated using a current-voltage formulation. While inequality constraints can be included in the optimization problem \eqref{eq:opf_eq}, they are omitted for simplicity in notation.

The I-V forecasting models are embedded directly into the equality constraints of the optimization problem as follows: 
\begin{subequations}
\label{eq:opf_forecast_eq}
    \begin{equation}
        \min_z f(z)
    \end{equation}
    \begin{equation}
       s.t. \sum_{i \in D_j} I_R^j (V_R^i,V_I^i,z) +g_{nn_i}^R(\bar{w},V_R,V_I,u) =0 
    \end{equation}
    \begin{equation}
       \sum_{i\in D_j} I_I^j (V_R^i,V_I^i,z)+g_{nn_i}^I(\bar{w},V_R,V_I,u) =0 \quad \forall i\in \mathcal{B} 
    \end{equation}
\end{subequations}

In this formulation, $g_{nn_i}^R$ and $g_{nn_i}^I$ are the real and imaginary current outputs of the trained I-V forecasting models in \eqref{eq:I-V_forecast_eq} for a device connected to bus $i$. These models are parameterized by a vector of ML model weights, $w_i$, and takes as input the real and imaginary components of the bus voltage ($V_R^j,V_I^j$) as well as exogenous features, $u$.

	By embedding the I-V forecasting models into the constraint sets, we capture the true current-voltage sensitivities of devices without assuming a predefined functional form. This allows the optimization algorithm to access more accurate first and second order gradients, leading to more reliable solutions.

\paragraph{Solving Hybrid ML-Physics Optimization}: To solve the constrained optimization problem in \eqref{eq:opf_forecast_eq}, we first construct the Lagrange equations as:

\begin{multline}
    \mathcal{L} =f(z)+ \\ \sum_{j\in\mathcal{B}} \lambda_j^R \left( \sum_{i \in D_j} I_R^i (V_R^j,V_I^j,z) +g_{nn_i}^R(\bar{w},V_R^j,V_I^j,u)
    \right) + \\ \lambda_j^I \left(\sum_{i \in D_j}I_I^i (V_R^j,V_I^j,z) +g_{nn_i}^I(\bar{w},V_R^j,V_I^j,u)
    \right)
\end{multline}
The stationary point of the Lagrange equation is determined by the following KKT conditions:

\begin{align}
    &\frac{\partial \mathcal{L}}{\partial z} = \frac{d f(z)}{d z} + \sum_{j\in\mathcal{B}} \lambda_j^R \left( \sum_{i \in D_j} \frac{d I_R^i}{d z}
    \right) + \lambda_j^I \left(\sum_{i \in D_j} \frac{d I_I^i}{d z}
    \right) \label{eq:kkt_1} \\
    &\frac{\partial \mathcal{L}}{\partial \lambda_j^R} = \sum_{i \in D_j} I_R^i (V_R^j,V_I^j,z) +g_{nn_i}^R(\bar{w},V_R^j,V_I^j,u)\label{eq:kkt_2} \\
     &\frac{\partial \mathcal{L}}{\partial \lambda_j^I} = \sum_{i \in D_j} I_I^i (V_R^j,V_I^j,z) +g_{nn_i}^I(\bar{w},V_R^j,V_I^j,u) \label{eq:kkt_3}\\
    &\frac{\partial \mathcal{L}}{\partial V_R^j} =  \sum_{j\in\mathcal{B}} \lambda_j^R \left( \sum_{i \in D_j} \frac{d I_R^i}{d V_R^j} + \frac{d g_{nn_i}^R}{dV_R^j}
    \right) + \lambda_j^I \left(\sum_{i \in D_j} \frac{d I_I^i}{d V_R^j} + \frac{d g_{nn_i}^I}{dV_R^j}
    \right) \label{eq:kkt_4}\\
   &\frac{\partial \mathcal{L}}{\partial V_I^j} =  \sum_{j\in\mathcal{B}} \lambda_j^R \left( \sum_{i \in D_j} \frac{d I_R^i}{d V_I^j} + \frac{d g_{nn_i}^R}{dV_I^j}
    \right) + \lambda_j^I \left(\sum_{i \in D_j} \frac{d I_I^i}{d V_I^j} + \frac{d g_{nn_i}^I}{dV_I^j}
    \right)   \label{eq:kkt_5}
\end{align}

The KKT conditions include evaluations physics-defined models (namely $I_R,I_I$), whose derivatives (such as ($dI_R/d V_R$  and $dI_I/dV_I$) can be determined analytically. Additionally, the KKT conditions include I-V forecasting models, $g_nn$, and partial derivatives with respect to bus voltages ($\frac{dg_{nn}^R}{dV_R },\frac{dg_{nn}^R}{dV_I} ,\frac{dg_{nn}^I}{dV_R},\frac{dg_{nn}^I}{dV_I}$). 

\paragraph{Solving KKT Conditions with I-V Forecast Models}
The KKT conditions in \eqref{eq:kkt_1}-\eqref{eq:kkt_5} construct the necessary first-order optimality conditions for solving the optimization problem in \eqref{eq:opf_forecast_eq}. However, including the ML forecasting model within the nonlinear set of equations requires integrating ML models within the numerical method. Our approach integrates the ML forecasting models within Newton-Raphson to solve the KKT conditions in \eqref{eq:kkt_1}-\eqref{eq:kkt_5}. Each NR step is expressed as:
\begin{equation}
\resizebox{ \columnwidth}{!} 
{$A = \begin{bmatrix}
\frac{d^2 f(z^k)}{dz^2} & \frac{d\Ir}{dz} & \frac{d \Ii}{dz} & 0 & 0\\
\frac{d\Ir}{dz} & 0 & 0 & \frac{d\Ir}{d\Vr} + \frac{d\gr}{d\Vr} & \frac{d\Ir}{d\Vi} + \frac{d\gr}{d\Vi} \\
\frac{d\Ii}{dz} & 0 & 0 & \frac{d\Ii}{d\Vr} + \frac{d\gi}{d\Vr} & \frac{d\Ii}{d\Vi} + \frac{d\gi}{d\Vi} \\
0 & \frac{d\Ir}{d\Vr} + \frac{d\gr}{d\Vr} & \frac{d\Ir}{d\Vi} + \frac{d\gr}{d\Vi} & 
\left( \frac{d^2\Ir}{d \Vr^2} + \frac{d^2 \gr}{d\Vr^2} \right)\lambda_R^T & 
\left( \frac{d^2\Ir}{d \Vr d\Vi} + \frac{d^2 \gr}{d\Vr d\Vi} \right)\lambda_I^T \\
0 & \frac{d\Ii}{d\Vr} + \frac{d\gi}{d\Vr} & \frac{d\Ii}{d\Vi} + \frac{d\gi}{d\Vi} & 
\left( \frac{d^2\Ii}{d \Vr d\Vi} + \frac{d^2 \gi}{d\Vr d\Vi} \right)\lambda_R^T & 
\left( \frac{d^2\Ii}{d\Vi^2} + \frac{d^2 \gi}{d\Vi^2} \right)\lambda_I^T
\end{bmatrix}$
}
\label{eq:kkt_NR_1}
\end{equation}

\begin{equation}
    B = 
\begin{bmatrix}
\frac{d f(z)}{d z} + \left( \frac{d \Ir^i}{d z} \right)^T \lambda^R + \left( \frac{d \Ii^i}{d z} \right)^T \lambda^I \\
\Ir^i + \gr(\bar{w}, \Vr, \Vi, u) \\
\Ii^i(\Vr, \Vi, z) + \gi(\bar{w}, \Vr, \Vi, u) \\
\left( \frac{d \Ir^i}{d \Vr} + \frac{d \gr}{d\Vr} \right)^T \lambda^R + 
\left( \frac{d \Ii}{d \Vr} + \frac{d \gi}{d\Vr} \right)^T \lambda^I \\
\left( \frac{d \Ir}{d \Vi} + \frac{d \gr}{d\Vi} \right)^T \lambda^R + 
\left( \frac{d \Ii}{d \Vi} + \frac{d \gi}{d\Vi} \right)^T \lambda^I 
\end{bmatrix}
\label{eq:kkt_NR_2}
\end{equation}

    \begin{equation}
        A 
\begin{bmatrix}
\Delta z^k \\
\Delta \lambda_R^k \\
\Delta \lambda_I^k \\
\Delta \Vr^k \\
\Delta \Vi^k
\end{bmatrix}
= -B
\label{eq:kkt_NR_3}
\end{equation}

\noindent where $A$ defined in \eqref{eq:kkt_NR_1} represents the Jacobian of the KKT conditions and $B$ defined in \eqref{eq:kkt_NR_2} is a vector of KKT conditions. The Newton-Raphson step for solving the KKT conditions is defined in \eqref{eq:kkt_NR_3}. $\Ir,\Ii,\Vr,\Vi$ are vectors representing the real and imaginary currents and voltages of all buses. Additionally, we model all forecasting models of devices as a vector $\gr$ and $\gi$ representing the real and imaginary components of the forecast.

In the Newton-Raphson step, we observe that the partials with respect to $\Ir,\Ii$ and $f(z)$ are derived analytically. On the other hand, to integrate the trained ML models within the NR step, we determine the partials, $\gr$ and $\gi$, through a combination of backpropagation and Hessian evaluations.  Specifically, the terms $\frac{d\gr}{d\Vr},\frac{d\gr}{d\Vi},\frac{d\gi}{d\Vr}, \frac{d\gi}{d\Vi}$  are computed using backpropagation of the ML model shown in Algorithm \ref{power_flow_alg}. The second-order derivatives, $\frac{d^2 \gr}{d\Vr^2},\frac{d^2\gr}{d\Vr d\Vi},\frac{d^2 \gi}{d\Vr d\Vi},\frac{d^2\gi}{d\Vi^2}$, are computed using the Hessian of the ML models as shown in Algorithm \ref{hessian_alg}. 

\begin{algorithm}
    \caption{Computing Hessians of I-V Forecast Models Using PyTorch  \cite{pytorch}}
    \label{hessian_alg}
    \textbf{Input: } $g_{nn}$($\cdot$), $\bar{w},u,V_R,V_I$
    \begin{algorithmic}[1]
 
    \STATE{ $[I_R,I_I ]=g_{nn} (\bar{w},V_R,V_I,u)$)}
    \STATE{ $g_{nn}.$backward(retain\_graph=$True$)}
    \STATE{ $[\frac{d^2 I_R}{d u^2}, \frac{d^2 I_R}{d V_R^2}, \frac{d^2 I_R}{d V_I^2}, \frac{d^2 I_R}{dV_R dV_I}]$=$I_R$.hessian.detach().clone()}
    \STATE{ $[\frac{d^2 I_I}{d u^2}, \frac{d^2 I_I}{d V_R^2}, \frac{d^2 I_I^2}{d V_I^2},\frac{d^2I_I}{dV_R dV_I}]$=$I_I$.hessian.detach().clone()}
  
    \RETURN $[\frac{d^2 I_R}{d u^2}, \frac{d^2 I_R}{d V_R^2}, \frac{d^2 I_R}{d V_I^2}, \frac{d^2 I_R}{dV_R dV_I},\frac{d^2 I_I}{d u^2}, \frac{d^2 I_I}{d V_R^2}, \frac{d^2 I_I^2}{d V_I^2},\frac{d^2I_I}{dV_R dV_I}]$
    \end{algorithmic}
    \end{algorithm}

The overall computation of the NR step then involves backpropagating and evaluating the Hessian of all ML models. This allows us to extract accurate sensitivities of the forecasted device learned through data for more accurate optimization solution. 

\section{Deriving Sensitivities of External Factors}

In the previous section, we integrated I–V forecast models into the optimization constraints of \eqref{eq:opf_forecast_eq} to improve solution accuracy. This integration also enables the computation of sensitivities of system-wide performance metrics with respect to exogenous features, $u$, such as weather or time-of-day, which influence the behavior of forecasted devices. These sensitivities are critical for understanding how exogenous factors impact grid performance and for enabling proactive, data-informed operational and planning decisions.

We define a system performance metric $g(x)$ to represent operational objectives for grid operators and planners, such as total active power generation from synchronous machines, constraint violations (infeasibility), or power losses across transmission lines. These metrics are functions of the system state variables, denoted $x$, which describe the operating condition of the grid. In the context of power flow, $x$ represent the real and imaginary components of bus voltages, as detailed in \cite{bromberg2015equivalent}. In the context of grid optimization, $x$ includes the optimization decision variables $z$, real and imaginary bus voltages and the corresponding real and imaginary parts of the dual variables, $\lambda_R$ and $\lambda_I$.

Our objective is to compute the sensitivity $\frac{dg(x)}{du}$, which measures how changes in these exogenous factors $u$ affect system metrics ,$g(x)$. These sensitivities offer insights that help grid operators identify the impact of exogenous factors and prioritize the most effective remedial actions. Additionally, they enable gradient-based optimization or control strategies that leverage these sensitivities to improve grid efficiency, measure risk, and find optimal solutions that ensure the system remains robust to a wide range of exogenous conditions.

The sensitivity of $g(x)$ to the exogenous factor, $u$, can be computed using a combination of the chain rule and backpropagation as:

\begin{equation}
\frac{dg(x)}{du} = \frac{dg(x)}{dx} \frac{dx}{dI_R}\frac{dI_R}{du} + \frac{dg(x)}{x}\frac{dx}{dI_I} \frac{dI_I}{du}
\label{eq:sensitivity_eq}
\end{equation}

Here, we observe two types of calculations required to evaluate $\frac{dg(x)}{du}$. The values $\frac{x)}{dI_R}$ and $\frac{(x)}{dI_I}$ represent the sensitivity of the system states to the real and imaginary currents of the forecasted devices. Then, the values $\frac{dI_R}{du}$ and $\frac{dI_I}{du}$ capture the sensitivity of each forecasted device with respect to exogenous factors, $u$.

To compute the first-order sensitivities $\frac{dg(x)}{dI_R}$ and $\frac{dg(x)}{dI_I}$, we study the linearized solution of the power flow and optimization problems described as:
\begin{equation}
Y x^* = J
\end{equation}

where $x^*$ is the solution of either the power flow or optimization problems. Here, $Y$ represents the Jacobian of power flow \eqref{eq:kcl_forecast_eq1}-\eqref{eq:kcl_forecast_eq2} or KKT conditions \eqref{eq:kkt_1}-\eqref{eq:kkt_5}, and the right-hand side vector $J$ captures the sum of device currents in power flow and optimization. The partials are then determined via the chain rule:

\begin{equation}
Y \frac{dx}{dI_R} = \frac{dJ}{dI_R}
\end{equation}

\begin{equation}
Y \frac{dx}{dI_I} = \frac{dJ}{dI_I}
\end{equation}

The sensitivities on the right, $\frac{dJ}{dI_R}$ and $\frac{dJ}{dI_I}$, are analytically computed, and for the I-V formulation are zeros except for a row of 1 representing the location of the forecasted device. Then, the sensitivities $\frac{dx}{dI_R}$ and $\frac{dx}{dI_I}$ are given by:

\begin{equation}
\frac{dx}{dI_R} = Y^{-1} \frac{dJ}{dI_R}
\label{eq:sensitivity_x_dIR}
\end{equation}

\begin{equation}
\frac{dx}{dI_I} = Y^{-1} \frac{dJ}{dI_I}
\label{eq:sensitivity_x_dII}
\end{equation}

where $Y^{-1}$ is pre-computed from the final Newton-Raphson step of power flow or optimization problem.

To evaluate \eqref{eq:sensitivity_eq}, the first-order sensitivities of the system state vector with respect to the forecasted device currents (computed in \eqref{eq:sensitivity_x_dIR}–\eqref{eq:sensitivity_x_dII}) are multiplied by $\frac{dI_R}{du}$ and $\frac{dI_I}{du}$, which are obtained via backpropagation through the I–V forecasting models with respect to the input features $u$. This backpropagation process is detailed in Algorithm \ref{power_flow_alg}.

Each term of \eqref{eq:sensitivity_eq} is computed independently and then combined to quantify the impact  of external factors on the system performance metric. Quantifying the sensitivity of grid performance to external factors like weather and time-of-day helps operators understand how changes in solar irradiance, temperature, or load patterns affect voltage profiles, power losses, and constraint violations. This enables more accurate forecasting and adaptive control strategies that maintain system reliability under varying conditions.

\section*{A. Using Sensitivities for System-Wide Optimization}

Our approach also enables us to design system-wide optimization problems where the decision variables are not direct physical setpoints but rather external factors that influence device behavior through machine learning-based forecasts. This opens the door to robust optimization strategies, where we can explore worst-case scenarios by identifying external conditions that pose the greatest risk to system performance and proactively design control strategies and policies that maintain grid stability.

This optimization can be represented by:

\begin{equation}
\min_u \, g(x)
\end{equation}

In this formulation, $x$ denotes the state variables of the nonlinear problem (i.e., real and imaginary voltages for power flow and dual variables for optimization). This optimization framework can be applied to quantify risk from external factors or serve as the inner problem within a bilevel robust optimization formulation.

With the sensitivities $\frac{dg}{du}$ computed, we can apply gradient-based optimization methods to adjust $u$ in order to improve the system-level objective $g(x)$ as:

\begin{equation}
u^{(k+1)} = u^{(k)} - \alpha \frac{dg(x)}{du}
\end{equation}

where $\alpha \in \mathbb{R}$ is a step-size for the gradient descent update, with the gradient $\frac{dg(x)}{du}$ computed through a combination of power flow sensitivities and backpropagation in \eqref{eq:sensitivity_eq}.

Optimizing over $u$ can enable robust operation under uncertainty, such as extreme weather or fluctuating loads. It also provides a framework for designing pricing mechanisms and can reveal which external factors have the greatest impact on performance, allowing grid operators to develop mitigation or curtailment strategies.

\section{Experiments}
Our approach offers three key benefits for power system operations: (1) provides accurate sensitivity of bus voltages for any device modeled through forecasting, (2) enables the computation of system performance sensitivities with respect to exogenous factors such as weather and time-of-day, and (3) supports system-level optimization aimed at improving grid robustness under real-world uncertainty. 

\begin{figure}
    \centering
    \includegraphics[width=0.6\linewidth]{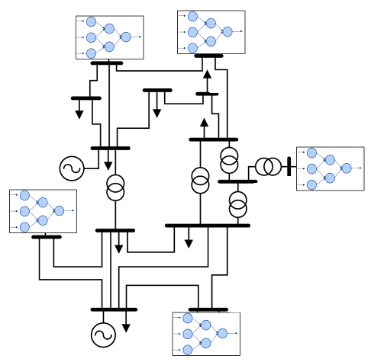}
    \caption{The current-voltage relation of renewable energy sources and loads are modeled by I-V forecasting models that are integrated into the optimization engine for 14-bus network.}
    \label{fig:14-bus-w-nn}
\end{figure}

To demonstrate the effectiveness of our method, we conduct experiments on a modified IEEE 14-bus network, where loads and renewable energy sources are modeled using LSTM-based forecasting. The LSTM model takes real and imaginary bus voltages, solar radiation, and ambient temperature as inputs from dataset in \cite{ramos2012modelling} to predict the behavior of renewable energy resources as well as loads. These forecast-driven models are placed at buses shown in Figure \ref{fig:14-bus-w-nn}, to capture realistic environmental variability.
Our goal is to demonstrate the capabilities enabled by integrating ML forecasts into power flow and optimization. This work establishes a foundation for future optimization methods to respond more effectively to real-world uncertainty including load and weather variations and determine their effect on the power grid. The main advantage of our approach is that it leverages ML models to characterize the local behavior of individual forecasted devices (such as loads and renewables), and integrates them within our grid optimization framework to assess their system-wide impacts. Studying these impacts enables operators and planners to better understand how localized, data-driven behaviors aggregate to influence global grid performance and helps to design more informed decisions in planning, control, and risk mitigation.

\subsection{Accurate Current-Voltage Sensitivities}
To demonstrate the accuracy of integrating the I-V forecasting model into power flow, we study the behavior of the 14-bus network where the loads represent a distribution network. These loads are modeled via (a) surrogate PQ model, (b) surrogate composite load, and (c) I-V forecasting model, whose hyperparameters are trained via simulated data. 

\begin{figure*}
\centering
\begin{subfigure}{.45\textwidth}
  \centering
  \includegraphics[width=.8\linewidth]{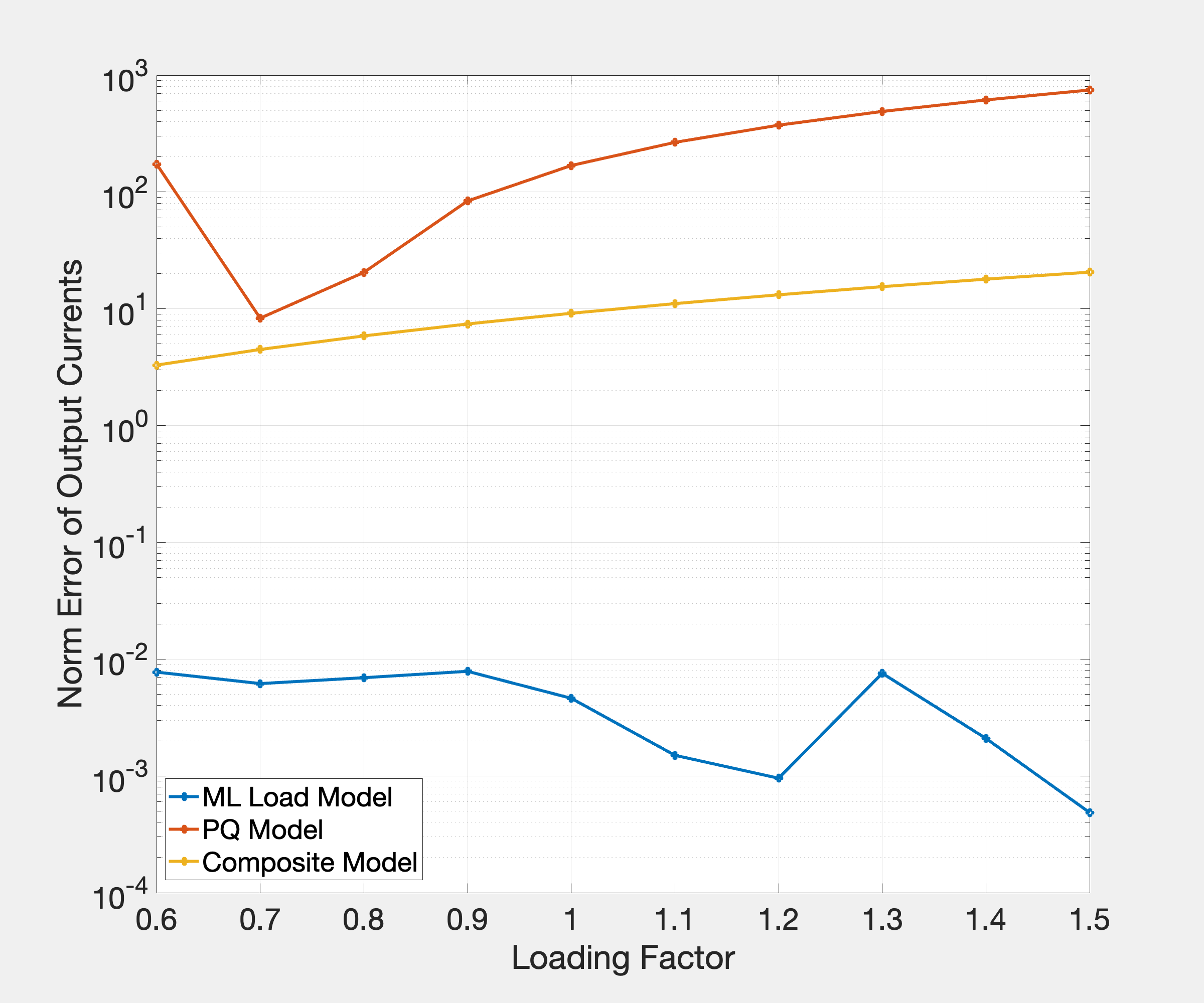}
  \caption{Output error compared to the ground truth distribution network response.}
  \label{fig:output_error}
\end{subfigure}%
\hfill
\begin{subfigure}{.45\textwidth}
  \centering
  \includegraphics[width=.8\linewidth]{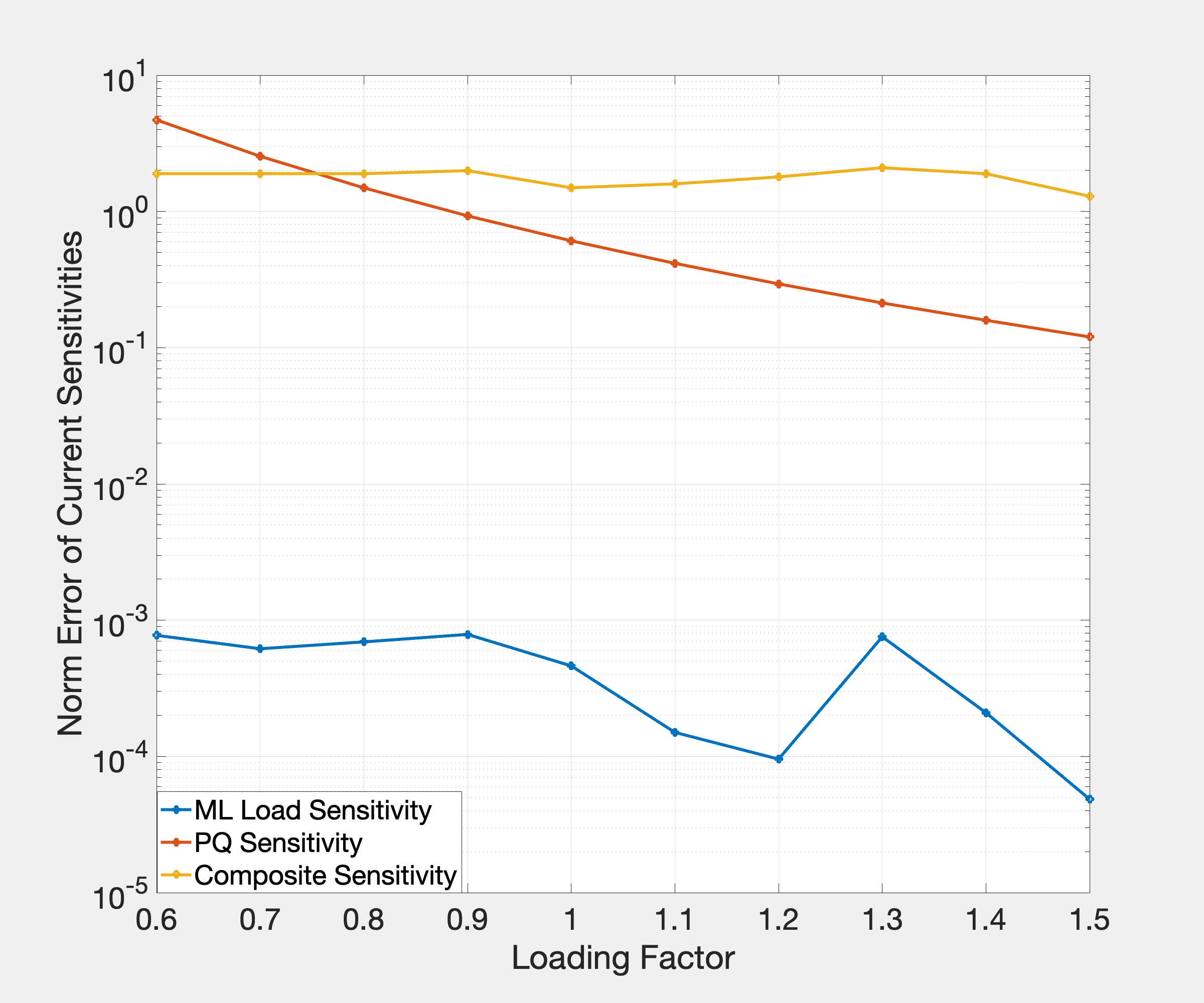}
  \caption{Sensitivity error compared to the ground truth distribution network sensitivity.}
  \label{fig:sensitivity_error}
\end{subfigure}
\caption{The behavior of the IEEE-8500 node distribution network is modeled by: (a) PQ load model, (b) composite load model, and (c) I-V forecasting model implemented using an LSTM network. For each model, parameters are fitted using a training dataset generated from voltage inputs ranging from 0.8 to 1.2 pu. The accuracy of the output predictions and the sensitivity of each model are then evaluated based on the network’s response.}
\label{fig:distribution_example}
\end{figure*}

Surrogate models macromodel the behavior of a complex distribution network as simple PQ or composite parameters. However, as shown in Figure \ref{fig:distribution_example}, this fails to capture the nonlinear current-voltage behavior and sensitivities of the network which leads to inaccurate simulations when there occur large variations in bus voltages. In contrast, our proposed I-V forecasting model is trained directly on simulated data and accurately captures the nonlinear behavior of the distribution network. This leads to more accurate power flow simulations under contingency scenarios as shown in Table \ref{tab:contingency_scenario}, where the I-V forecasting model precisely predicts the nonlinear behavior of individual devices. 

\begin{table}[]
    \centering
        \caption{The network in Figure \ref{fig:14-bus-w-nn} undergoes a contingency and increased loading factor. Renewable energy sources and loads are modeled by (a) PQ models, (b) composite models, and (c) I-V Forecast Models. The minimum and maximum voltages of the simulation results are listed, as well as the percentage error with the ground truth simulation.}
    \begin{tabular}{|c|p{1.8cm}|p{1.8cm}|p{1.2cm}|}
    \hline
        Load Model & Minimum Voltage (pu) & Maximum Voltage (pu) & State Error (\%) \\
        \hline 
        PQ Model & 0.96 & 1.22 &0 \\
        \hline 
Composite Model & 0.96 & 1.22 &0 \\
\hline 
I-V Forecast Model & 0.96 & 1.22 &0 \\
\hline
    \end{tabular}
    \label{tab:contingency_scenario}
\end{table}

\subsection{Accurate Sensitivities to Exogenous Factors}
\begin{figure*}
\centering
\begin{subfigure}{.45\textwidth}
  \centering
  \includegraphics[width=.8\linewidth]{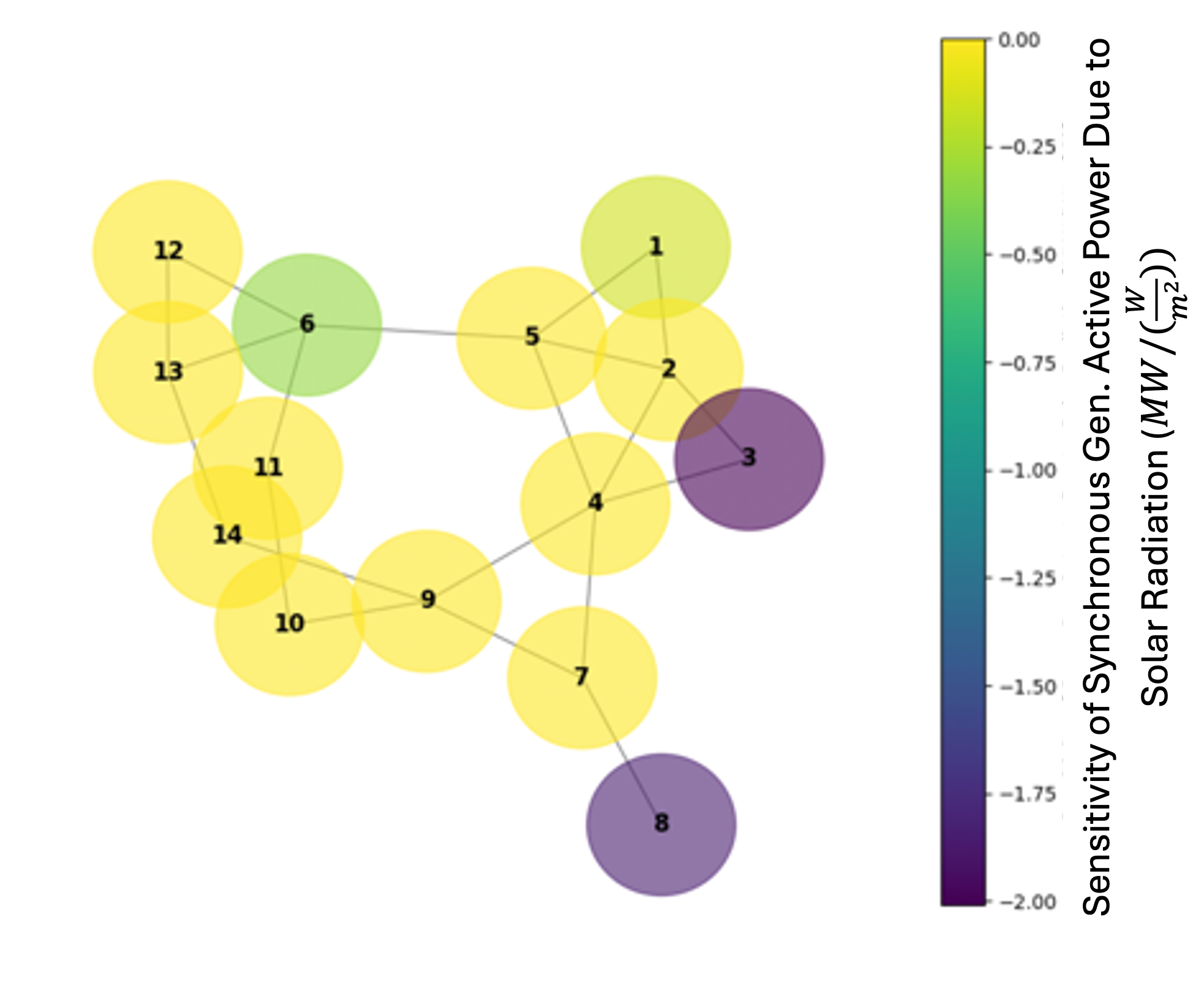}

  \label{fig:sensitivity_P}
\end{subfigure}%
\begin{subfigure}{.45\textwidth}
  \centering
  \includegraphics[width=.8\linewidth]{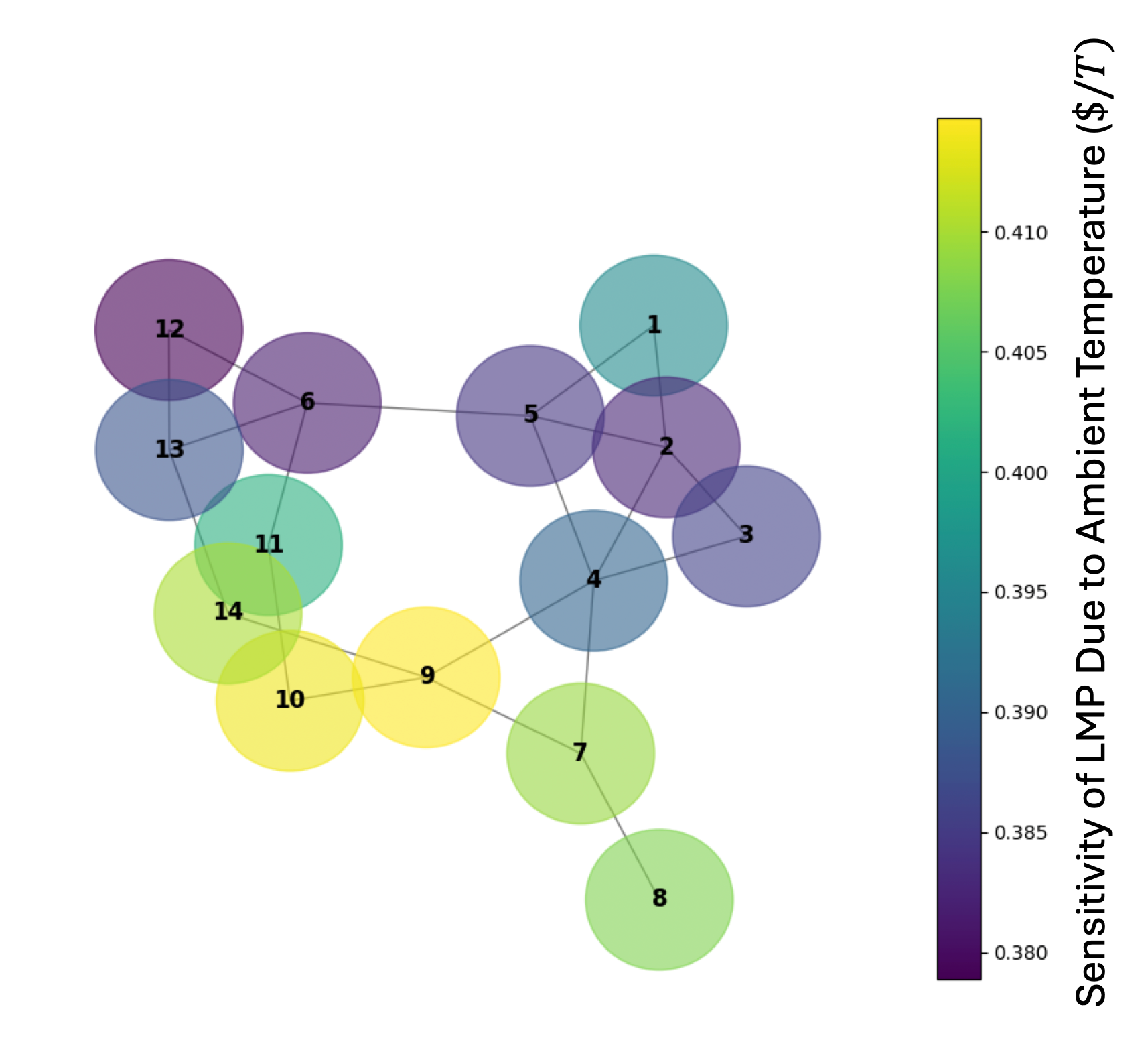}

  \label{fig:sensitivity_LMP}
\end{subfigure}
\caption{System performance impacts from environmental changes are computed using sensitivities from physics-based and DNN-based forecasting models. The left shows active power changes of synchronous generators due to solar radiation, and the right shows effect on Locational Margin due to ambient temperature.}
\label{fig:external_sensitivity_figures}
\end{figure*}

Integrating ML models into the power flow framework not only improves the accuracy of voltages sensitivities but also enables access to sensitivities with respect to exogenous factors. In this experiment, we study the sensitivity of system performance metrics with respect to solar radiation and ambient temperature.
Figure \ref{fig:external_sensitivity_figures}(a) illustrates the sensitivity of active power generation of synchronous machines to solar radiation levels. This sensitivity emerges from the interaction between forecasted renewable generation and power flow constraints. The I-V forecasting model captures how increased solar irradiance impacts local renewable generation, which is then combined with the sensitivity of the transmission grid to measure the sensitivity to synchronous machines. 
In another study, we assess how ambient temperature affects locational marginal prices (LMPs) as shown in Figure \ref{fig:external_sensitivity_figures}(b). Our methodology captures the impact of ambient temperatures on local load behaviors and combines their effects on marginal cost of all buses in the grid. This allows operators to understand the impact that exogenous factors have on pricing.
To validate the accuracy of these sensitivities, we compare our results with direct perturbations in the power flow simulations. As shown in Table \ref{tab:sensitivity_perturbations}, the first-order sensitivities computed by our method closely match changes resulting from 10\% variations in exogenous inputs. These results confirm that our ML-integrated power flow reliably captures the influence of exogenous conditions on grid behavior and can provide insights for operations and planning.

\begin{table}[h]
    \centering
        \caption{Accuracy of Sensitivity Calculations Compared to Power Flow Simulations with 10\% Perturbation in Exogenous Features}
    \begin{tabular}{|p{2cm}|p{2cm}|p{3cm}|}
        \hline
         & Sensitivity of $P_G$ with respect to solar radiation & Sensitivity of LMP with respect to ambient temperature \\
         \hline
        Mean Percentage Error (\%) & 0.1 & 2.5 \\
        \hline
    \end{tabular}
    \label{tab:sensitivity_perturbations}
\end{table}

\subsection{Using Sensitivities for Robust Optimization}

We leverage the computed sensitivities as part of a larger grid optimization framework aimed at designing system dispatch strategies that are robust under significant fluctuations in ambient temperature. By incorporating the sensitivities from Figure \ref{fig:external_sensitivity_figures} into the optimization problem defined in Equation \eqref{eq:sensitivity_eq}, we construct a bilevel optimization problem where exogenous weather variability is treated as an adversarial input:

\begin{equation}
\min_{z} \max_{u} \; g(x)
\end{equation}

\noindent
subject to
\begin{align}
\sum_{i \in D_j} \left[ I_R^j(V_R^i, V_I^i, z) + g_{nn,i}^R(w_i, V_R^j, V_I^j, u_i) \right] &= 0 \\
\sum_{i \in D_j} \left[ I_I^j(V_R^i, V_I^i, z) + g_{nn,i}^I(w_i, V_R^j, V_I^j, u_i) \right] &= 0 \quad \forall j \in B
\end{align}

This formulation is solved using an attacker-defender approach, as outlined in \cite{agarwal2022employing}. In this strategy, the attacker—represented by the inner maximization—aims to increase system infeasibilities and operational costs by perturbing temperature profiles within realistic bounds. This models worst-case weather scenarios. The inner maximization problem is defined as:

\begin{equation}
\max_{u} \; g(x)
\end{equation}

\noindent
which is solved using a gradient ascent method with the iterative update rule:

\begin{equation}
u^{(k+1)} = u^{(k)} + \alpha \frac{d g(x, z^*)}{d u}
\end{equation}

\noindent
where $\alpha \in (0, 1]$ is the step size. We iteratively solve the attack stage using gradients derived through our method in Equation \eqref{eq:sensitivity_eq} to obtain the worst-case temperature profile $u^*$ that maximizes infeasibilities for a given dispatch setpoint $z^*$.

The defender, represented by the outer minimization, then computes an optimal dispatch strategy that minimizes the impact of these adversarial conditions. The defender problem is:

\begin{equation}
\min_{z} \; g(x)
\end{equation}

\noindent
subject to
\begin{align}
\sum_{i \in D_j} \left[ I_R^j(V_R^i, V_I^i, z) + g_{nn,i}^R(w_i, V_R^j, V_I^j, u^*) \right] &= 0 \\
\sum_{i \in D_j} \left[ I_I^j(V_R^i, V_I^i, z) + g_{nn,i}^I(w_i, V_R^j, V_I^j, u^*) \right] &= 0 \quad \forall j \in B
\end{align}

\noindent
This interplay between the attacker and defender results in a system dispatch that is inherently robust to extreme operating conditions.

We demonstrate the effectiveness of this method on a 14-bus test network. The results show that the dispatch from the bilevel optimization significantly reduces network constraint violations (i.e., KCL) and cost of secondary generation under large temperature variations. This illustrates the critical role that ML-derived sensitivities play in improving grid optimization, particularly in designing robust dispatch strategies.

\begin{figure}
    \centering
    \includegraphics[width=0.8\linewidth]{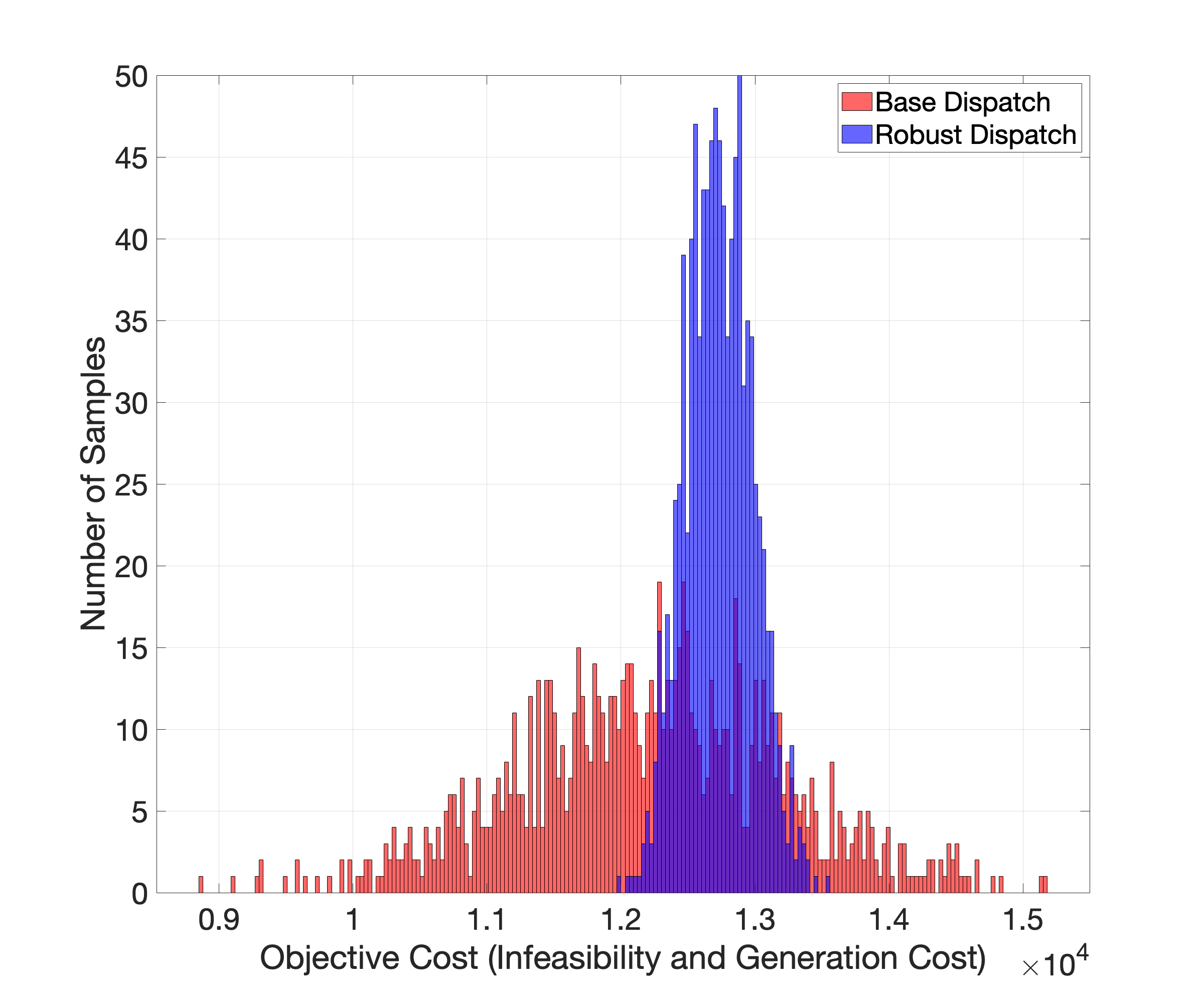}
    \caption{A stochastic bilevel-optimization framework uses sensitivities from I-V forecasting models to produce a dispatch that is optimized for robustness against temperature changes. A Monte-Carlo analysis for infeasibilities and cost of generation over a distribution of temperatures shows improvement for the optimized system compared to base dispatch.}
    \label{fig:robust_result}
\end{figure}
\section{Conclusion}

	We present a framework that integrates ML-based forecasting models directly into power flow and grid optimization. This eliminates the need for inaccurate surrogate models and captures the sensitivity of bus voltages as well as the impact of external factors like weather and time-of-day on system-wide performances. We derive these sensitivities by combining backpropagation from ML forecasts with gradients of physics-based grid models. Our results demonstrate that this approach not only improves accuracy of sensitivity calculations and power flow simulations but also enables the design of robust grid strategies capable of withstanding large changes in external factors. The experiments presented in this work highlight how the sensitivities of system-wide metrics with respect to external factors can be used to inform system-level decisions, assess external risks, and design control strategies that respond to real-world variability. This points to new directions for research surrounding the power grid in large-scale optimization, robust planning, and market design. 

\bibliographystyle{IEEEtran}

\bibliography{main} 

\begin{thebibliography}{10}
\providecommand{\url}[1]{#1}
\csname url@samestyle\endcsname
\providecommand{\newblock}{\relax}
\providecommand{\bibinfo}[2]{#2}
\providecommand{\BIBentrySTDinterwordspacing}{\spaceskip=0pt\relax}
\providecommand{\BIBentryALTinterwordstretchfactor}{4}
\providecommand{\BIBentryALTinterwordspacing}{\spaceskip=\fontdimen2\font plus
\BIBentryALTinterwordstretchfactor\fontdimen3\font minus \fontdimen4\font\relax}
\providecommand{\BIBforeignlanguage}[2]{{%
\expandafter\ifx\csname l@#1\endcsname\relax
\typeout{** WARNING: IEEEtran.bst: No hyphenation pattern has been}%
\typeout{** loaded for the language `#1'. Using the pattern for}%
\typeout{** the default language instead.}%
\else
\language=\csname l@#1\endcsname
\fi
#2}}
\providecommand{\BIBdecl}{\relax}
\BIBdecl

\bibitem{hong2015weather}
T.~Hong, P.~Wang, and L.~White, ``Weather station selection for electric load forecasting,'' \emph{International Journal of Forecasting}, vol.~31, no.~2, pp. 286--295, 2015.

\bibitem{weron2014electricity}
R.~Weron, ``Electricity price forecasting: A review of the state-of-the-art with a look into the future,'' \emph{International journal of forecasting}, vol.~30, no.~4, pp. 1030--1081, 2014.

\bibitem{hyndman2018forecasting}
R.~J. Hyndman and G.~Athanasopoulos, \emph{Forecasting: principles and practice}.\hskip 1em plus 0.5em minus 0.4em\relax OTexts, 2018.

\bibitem{taylor2003short}
J.~W. Taylor, ``Short-term electricity demand forecasting using double seasonal exponential smoothing,'' \emph{Journal of the Operational Research Society}, vol.~54, no.~8, pp. 799--805, 2003.

\bibitem{hippert2002neural}
H.~S. Hippert, C.~E. Pedreira, and R.~C. Souza, ``Neural networks for short-term load forecasting: A review and evaluation,'' \emph{IEEE Transactions on power systems}, vol.~16, no.~1, pp. 44--55, 2002.

\bibitem{arora2021remodelling}
P.~Arora, A.~Khosravi, B.~K. Panigrahi, and P.~N. Suganthan, ``Remodelling state-space prediction with deep neural networks for probabilistic load forecasting,'' \emph{IEEE Transactions on Emerging Topics in Computational Intelligence}, vol.~6, no.~3, pp. 628--637, 2021.

\bibitem{wazirali2023state}
R.~Wazirali, E.~Yaghoubi, M.~S.~S. Abujazar, R.~Ahmad, and A.~H. Vakili, ``State-of-the-art review on energy and load forecasting in microgrids using artificial neural networks, machine learning, and deep learning techniques,'' \emph{Electric power systems research}, vol. 225, p. 109792, 2023.

\bibitem{eskandari2021convolutional}
H.~Eskandari, M.~Imani, and M.~P. Moghaddam, ``Convolutional and recurrent neural network based model for short-term load forecasting,'' \emph{Electric Power Systems Research}, vol. 195, p. 107173, 2021.

\bibitem{wu2021support}
J.~Wu, Y.-G. Wang, Y.-C. Tian, K.~Burrage, and T.~Cao, ``Support vector regression with asymmetric loss for optimal electric load forecasting,'' \emph{Energy}, vol. 223, p. 119969, 2021.

\bibitem{fan2022applications}
G.-F. Fan, L.-Z. Zhang, M.~Yu, W.-C. Hong, and S.-Q. Dong, ``Applications of random forest in multivariable response surface for short-term load forecasting,'' \emph{International Journal of Electrical Power \& Energy Systems}, vol. 139, p. 108073, 2022.

\bibitem{matrenin2022medium}
P.~Matrenin, M.~Safaraliev, S.~Dmitriev, S.~Kokin, A.~Ghulomzoda, and S.~Mitrofanov, ``Medium-term load forecasting in isolated power systems based on ensemble machine learning models,'' \emph{Energy Reports}, vol.~8, pp. 612--618, 2022.

\bibitem{sobolewski2023gradient}
R.~A. Sobolewski, M.~Tchakorom, and R.~Couturier, ``Gradient boosting-based approach for short-and medium-term wind turbine output power prediction,'' \emph{Renewable Energy}, vol. 203, pp. 142--160, 2023.

\bibitem{dubey2021study}
A.~K. Dubey, A.~Kumar, V.~Garc{\'\i}a-D{\'\i}az, A.~K. Sharma, and K.~Kanhaiya, ``Study and analysis of sarima and lstm in forecasting time series data,'' \emph{Sustainable Energy Technologies and Assessments}, vol.~47, p. 101474, 2021.

\bibitem{huang2022time}
X.~Huang, Q.~Li, Y.~Tai, Z.~Chen, J.~Liu, J.~Shi, and W.~Liu, ``Time series forecasting for hourly photovoltaic power using conditional generative adversarial network and bi-lstm,'' \emph{Energy}, vol. 246, p. 123403, 2022.

\bibitem{aseeri2023effective}
A.~O. Aseeri, ``Effective rnn-based forecasting methodology design for improving short-term power load forecasts: Application to large-scale power-grid time series,'' \emph{Journal of Computational Science}, vol.~68, p. 101984, 2023.

\bibitem{bromberg2015equivalent}
D.~M. Bromberg, M.~Jereminov, X.~Li, G.~Hug, and L.~Pileggi, ``An equivalent circuit formulation of the power flow problem with current and voltage state variables,'' in \emph{2015 IEEE Eindhoven PowerTech}.\hskip 1em plus 0.5em minus 0.4em\relax IEEE, 2015, pp. 1--6.

\bibitem{pytorch}
A.~Paszke, S.~Gross, S.~Chintala, G.~Chanan, E.~Yang, Z.~DeVito, Z.~Lin, A.~Desmaison, L.~Antiga, and A.~Lerer, ``Automatic differentiation in pytorch,'' in \emph{NIPS-W}, 2017.

\bibitem{ramos2012modelling}
S.~Ramos, M.~Silva, F.~Fernandes, and Z.~Vale, ``Modelling real solar cell using pscad/matlab,'' 2012.

\bibitem{agarwal2022employing}
A.~Agarwal, P.~L. Donti, J.~Z. Kolter, and L.~Pileggi, ``Employing adversarial robustness techniques for large-scale stochastic optimal power flow,'' \emph{Electric Power Systems Research}, vol. 212, p. 108497, 2022.

\end{thebibliography}

\end{document}